\documentclass[aps,prl,twocolumn,showpacs]{revtex4-1}
\usepackage{epsfig}
\usepackage{graphicx}
\usepackage{amsmath}
\usepackage{comment}
\usepackage{bm}

\allowdisplaybreaks

\begin{document}

\DeclareGraphicsExtensions{.eps,.EPS,.pdf}

\title{Measuring correlations from the collective spin fluctuations of a large ensemble of lattice-trapped dipolar spin-3 atoms}
\author{Youssef Aziz Alaoui$^{1,2}$, Bihui Zhu$^{3}$, Sean Robert Muleady$^{4,5}$, William Dubosclard$^{1,2}$, Tommaso Roscilde$^{6}$, Ana Maria Rey$^{4,5}$, Bruno Laburthe-Tolra$^{2,1}$, and Laurent Vernac$^{1,2}$ }

\affiliation{$^{1}$\,Universit\'e Paris 13, Laboratoire de Physique des Lasers, F-93430, Villetaneuse, France\\
$^{2}$\,CNRS, UMR 7538, LPL, F-93430, Villetaneuse, France\\
$^{3}$\,Homer L. Dodge Department of Physics and Astronomy, The University of Oklahoma, Norman, Oklahoma 73019, USA \&\\
 Center for Quantum Research and Technology, The University of Oklahoma, Norman, Oklahoma 73019, USA\\
$^{4}$\,JILA, NIST and Department of Physics, University of Colorado, Boulder, USA\\
$^{5}$\,Center for Theory of Quantum Matter, University of Colorado, Boulder, CO 80309, USA\\
$^{6}$\, Univ Lyon, Ens de Lyon, CNRS, Laboratoire de Physique, F-69342 Lyon, France
}

\begin{abstract}

We perform collective spin measurements to study the buildup of two-body correlations between $\approx10^4$  spin $s=3$  chromium atoms pinned in a 3D optical lattice. The spins interact via long range and anisotropic dipolar interactions. From the fluctuations of total magnetization, measured at the standard quantum limit, we estimate the dynamical growth of the connected pairwise correlations associated with magnetization. The quantum nature of the correlations is assessed by comparisons with short and long time expansions, and numerical simulations. Our work shows that measuring fluctuations of spin populations provides new ways to characterize correlations in quantum many-body systems, for $s>1/2$ spins.

\end{abstract}
\date{\today}
\maketitle

Characterizing quantum correlations between different parts of a system is of fundamental importance for the development of quantum technologies and the study of complex quantum systems.  Quantum correlations are at the heart of the most peculiar effects predicted by quantum mechanics, such as entanglement, EPR steering \cite{Horodecki2009,Cavalcanti2016,Pezze2018}, or Bell nonlocality \cite{Brunner2014}, which all give advantage for different quantum information or metrological tasks. Quantum correlations can even arise for non-entangled states \cite{Ollivier2001,Luo2008}, where they can still constitute an interesting resource \cite{Bera2017}. Systems made of $s>1/2$ particles pinned in optical lattices are particularly interesting for such applications, as their Hilbert space, enlarged with respect to qubit ($s=1/2$) systems, offers new possibilities for quantum information processing \cite{Wang2020}.

Quantum correlations should appear in generic quantum systems \cite{Ferraro2010}, but proving their inherent  quantum nature is an experimental challenge, which requires the measurement of non-commuting operators. As full state-tomography scales exponentially with the number of constituents \cite{Kaufman2016} and thus becomes impossible in large ensembles, it is of crucial importance   to develop new protocols to infer correlations from partial measurements such as bipartite or collective measurements. The latter have been successful in demonstrating entanglement \cite{Pezze2018}, steering\cite{Fade2018,Kunkel2018,Lange2018}, or nonlocality \cite{Schmied2016}, in experimental platforms dealing with effective two-level systems, for which entanglement  witnesses have a simpler structure compared to particles with $s>1/2$  \cite{Sorensen2001, Vitagliano2011, Toth2014}. Extensions to $s=1$ systems in spinor Bose-Einstein condensates have demonstrated  number squeezing in  pair creation processes via spin-mixing collisions \cite{Pezze2018,Bookjans2011,Gross2011,Lucke2011,Qu2020},  SU(1,1) interferometry\cite{Linnemann2016}, and entangled fragmented phases \cite{Evrard2021}. These systems nevertheless operated in the regime where the single-mode approximation is valid \cite{Law1998}, which enormously simplifies the quantum dynamics.

\begin{figure}
\centering
\includegraphics[width= 3.5 in]{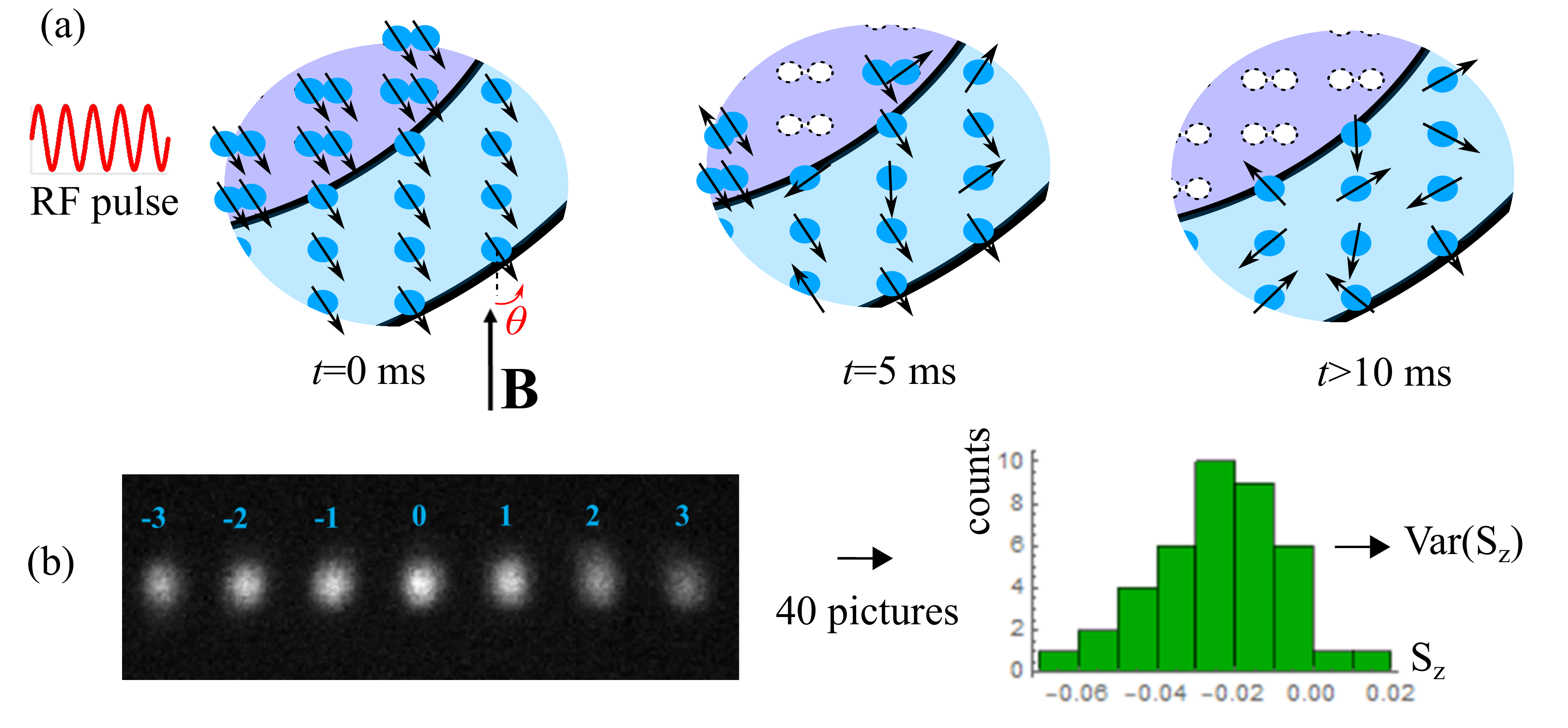}
\caption{Principle of the experiment. a) The cartoons zoom over a small region of the Mott insulating distribution with doubly (core) and simply occupied sites (shell). Spin 3 chromium atoms are excited  at $t=0$ by a RF pulse, with 5 cycles at the Larmor period set by the external magnetic field $\textbf{B}$. The spin directions then makes an angle $\theta$ (set to  $\pi/2$ in this work) with respect to $\textbf{B}$, which triggers spin dynamics. Correlations develop between spins, while doubly occupied sites get empty. b) Stern Gerlach separation provides  measurement of the fractional spin populations $p_{ms}(t)$, through fluorescence imaging, at a given dynamic time $t$ . Repeating the experiment allows us to compute the variance of the magnetization, and hence the correlator $C_z$ of Eq.(\ref{Cz}).}
\label{fig1}
\end{figure}

In this work we measure for the first time  two-body correlations in  a macroscopic array of  spin-3 chromium atoms   pinned in a 3D optical lattice and coupled via long-range and anisotropic magnetic dipolar interactions. Prior experiments  measuring   out of equilibrium spin dynamics in these arrays   demonstrated   compatibility with the growth of quantum correlations \cite{Lepoutre2019,Patscheider2020} and their   approach to  quantum thermalization \cite{Lepoutre2019}. Here, we make use of the large atomic spin to obtain a direct measurement of two-body correlations. Specifically, after  triggering  out-of equilibrium spin dynamics, we   acquire statistics on the $2s+1=7$ spin populations, and quantify the growth of inter-atomic spin correlations  by analyzing the statistical fluctuations of the collective spin component along the external magnetic field, i.e. the magnetization.
 The quantum nature of the correlations we measure is  validated by agreement with exact  short time expansions, with a  high temperature series expansion applied to the asymptotic quantum thermalized state  at long time, and with simulations of the full quantum dynamics via advanced phase-space numerical methods.

We consider a system  of $N$ spin  $s>1/2$ particles.  We define   $\hat s_z^i$  as the  z-component of the spin of the  $i^{th}$ particle. The correlator $ C_z$ we aim to measure is:
\begin{equation}
 C_z=\sum_{i\neq j}^{N}\left(\left \langle\hat{s}_z^i\hat{s}_z^j\right \rangle-\left \langle\hat{s}_z^i\right \rangle\left \langle\hat{s}_z^j\right \rangle\right)={\rm{Var}}(\hat S_z)-\Sigma_z
\label{Cz}
\end{equation}
with ${\rm{Var}}(\hat S_z)= \langle\hat S_z^2 \rangle- \langle\hat S_z \rangle^2$ the variance of the collective spin component $\hat S_{z}=\sum_{i=1}^{N}\hat s_z^i$, and $\Sigma_z$ the sum of individual variances, $\Sigma_z=\sum_{i=1}^{N}\Big(\left \langle(\hat s_z^i)^2\right \rangle-\left \langle\hat s_z^i\right \rangle^2\Big)$. $\Sigma_z$ accounts for intraparticle  correlations, which are only non-trivial  for  $s>1/2$, as $\left \langle ( {\hat s}_z^i)^2\right \rangle=1/4$ if $s=1/2$. The interparticle correlations are accounted for by the two-body correlator $C_z$.

In this work, we independently determine ${\rm{Var}}(\hat S_z)$ and $\Sigma_z$ from collective measurements in order to show that $C_z$ departs from 0. Measurement of ${\rm Var}(\hat S_z)$ requires the experiment to be repeated many times to acquire adequate statistics. Measurement of $\Sigma_z$ is straightforward in the case of an homogeneous system, comprising singly-occupied lattice sites, referred as singlons in the following. Indeed for singlons,  $\left \langle(\hat{s}_z^i)^2\right \rangle=\sum_{m_s}p_{m_s}^{(i)} m_s^2$, with $p_{m_s}^{(i)}$ the probability that the site $i$, uniquely populated by the $i$-th spin, is in the $m_s$ spin state ($-3\leq m_s\leq 3$ , $\sum_{m_s}p_{m_s}^{(i)}=1$) so that  $\sum_i\left \langle(\hat{s}_z^i)^2\right \rangle=N\sum_{m_s}p_{m_s}m_s^2$  even without homogeneity. Homogeneity ensures that $p_{m_s}^{(i)}=p_{m_s}$ are independent of site $i$, so that   $\left\langle\hat{s}_z^i\right \rangle^2 =(\sum_{m_s}p_{m_s}m_s)^2$; therefore
\begin{equation}
\frac{\Sigma_z}{N}=\sum_{m_s}p_{m_s}m_s^2-\left(\sum_{m_s}p_{m_s}m_s\right)^2
\label{Sigmaz}
\end{equation}
We show in Ref.~\cite{SuppMat} that the experimental magnetic inhomogeneities lead to negligible deviations from Eq.(\ref{Sigmaz}). Besides, inhomogeneities of the lattice potential are below $2.5 \% $, and therefore have a negligible effect on spin dynamics in the Mott regime \cite{petrapra2019}. Finally, we show in Ref.~\cite{SuppMat} that Eq.(\ref{Sigmaz}) still holds for doubly occupied sites (doublons) which are populated at the beginning of dynamics (see below). This is due to the fact that the spin of each pair of particles in a doubly-occupied site is well-defined at all times. Therefore, $\Sigma_z$ is given by Eq.(\ref{Sigmaz}) for the whole dynamics, and measurement of $p_{m_s}=N_{m_s}/N$, with $N_{m_s}$ the total number of atoms in spin state $m_s$, yields $\Sigma_z$.

{\it  Experimental setup}. The starting point of our experiments is a spin-polarized $^{52}$Cr Bose-Einstein Condensate (BEC) produced in a crossed dipole trap, with typically $15000$ atoms polarized in the minimal Zeeman energy state $m_s=-3$. We load the $^{52}$Cr BEC in a 3D optical lattice deep into the Mott insulator regime. The lattice implemented with five lasers at $\lambda_L=532$ nm is described in \cite{Lepoutre2019}. The total lattice depth is equal to 60 recoil energy at $\lambda_L$. We estimate the tunneling time to be $\simeq 20$ ms. We obtain a core of doublons comprising $\simeq50\%$ of the atoms, surrounded by a shell of singlons.

As shown in Fig. \ref{fig1}(a), we trigger spin dynamics by rotating all spins with the use of a Radio Frequency (RF) $\pi/2$ pulse. After the pulse all spins are oriented  orthogonal to the external magnetic field. The Larmor frequency $f_L=g_L\mu_B B_0/\hbar$ (with $g_L\simeq2$ the Land\'e factor, $\mu_B$ the Bohr magneton, and $B_0=0.75$ Gauss the amplitude of the magnetic field) is $f_L\simeq2.1$ MHz. The RF frequency $f_{RF}$ is set at resonance, and fluctuations of the detuning $(f_L-f_{RF})\simeq1$ kHz are small compared to the RF Rabi frequency $f_R$, thanks to  the use of a 30 Watt RF amplifier. In practice the RF pulse has a duration of exactly 5 Larmor periods, with $f_R=\frac{1}{5}\frac{f_L}{4}=105$ kHz; the  $\bar{\theta} =\pi/2$   pulse is set to have  an identical initial phase  at each realization. Fluctuations of the rotation angle $\theta$ are  estimated to have  a standard deviation of $\sigma_{\theta}\simeq2.5\times10^{-3}$ rad (see below). After the initial state preparation with the RF pulse, spins interact via  magnetic dipolar interactions  in the optical lattice for a duration $t$. We then adiabatically ramp down the optical lattice,  and proceed to measurements.

{\it  Theoretical model}
Dipolar interactions between singlons during the dark time evolution  are described by the effective dipolar Hamiltonian $\hat H_{\rm dd}$, which is a XXZ spin model Hamiltonian:
\begin{equation}
\hat H_{\rm dd}=\sum_{i> j}^{N} V_{ij} \left[ \hat s_z^i \hat s_z^j -\frac{1}{2} \left( \hat s_x^i \hat s_x^j + \hat s_y^i \hat s_y^j \right) \right]
\label{secular}
\end{equation}
with $V_{ij}=V_{dd} \left( \frac{1-3 \cos ^2 \theta _{ij}}{r_{ij}^3}\right)$, $V_{dd}= \frac{\mu_0 (g_L \mu_B)^2}{4 \pi}$, and $\mu_0$ the magnetic permeability of vacuum. The sum runs over all pairs of particles ($i$,$j$),   $r_{ij}$ is their  corresponding distance, $\theta_{ij}$ the angle between their inter-atomic axis and the external magnetic field, ${\bf{\hat s}}_i=\{\hat s_x^i,\hat s_y^i,\hat s_z^i \}$ are spin-3 angular momentum operators  for  atom $i$. The shortest intersite distance $r_{min}=268$ nm in our lattice \cite{Lepoutre2019} corresponds to a dipolar coupling $V_{dd}/r_{min}^3\simeq h \times 3$ Hz.

\begin{figure}
\centering
\includegraphics[width= 3.4 in]{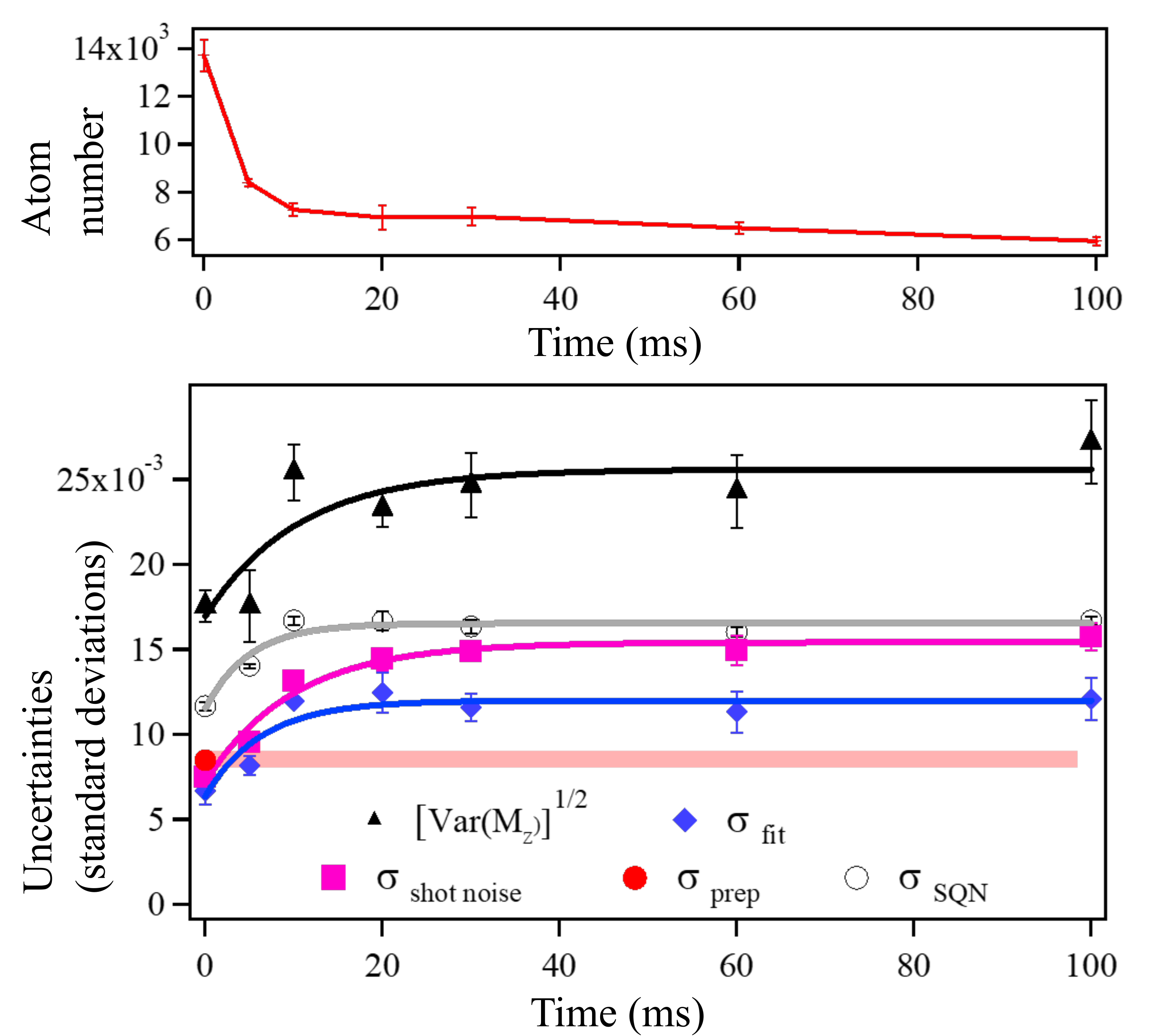}
\caption{Evolution of the atom number $N(t)$ (Top), and of the fluctuations measured in the experiment (Bottom): we show the standard deviations of the normalized magnetization $M_z$ and of the technical noises featured in Eq.(\ref{sumsigmaMz}): $\sigma_{{\rm{prep}}}$ (shaded area; preparation of the sample), $ \sigma_{{\rm{fit}}}$ (fitting uncertainties), and $\sigma_{{\rm{shot  noise}}}$ (fluorescence imaging). The quantum projection noise $\sqrt{\frac{3}{2N(t)}}=\sigma_{{\rm{SQN}}}$ is shown for comparison. Lines are guides to the eye. Error bars evaluated from statistics correspond to two standard deviations.}
\label{fig2}
\end{figure}

Given the strong contact interactions that favor spin alignment \cite{Lepoutre2018PRA,Lepoutre2018SCM} and the fully polarized initial state, the same Hamiltonian can be used to describe the  dynamics of doubly occupied sites (doublons)  just by  replacing ${\bf{\hat s}}_i$ by a  spin-6 angular momentum operator at the corresponding site \cite{petrapra2019}. Nevertheless, as soon as the spin excitation is performed, doublons start to leave the trap, due to dipolar relaxation \cite{depaz2013a}, see Fig.\ref{fig2}: for $0<t<10$ ms the spin system comprises both singlons and doublons, but only singlons remain for $t>10$ ms and losses become negligible. This is why we run simulations for singlons only, which allows for quantitative comparison with the experiment except at short time.

As shown  in previous work \cite{Lepoutre2019}, we need to include  the one-body term $\hat{H}_Q=B_{\rm Q}\sum_i^N (\hat s_z^i)^2$ accounting for  light shifts created by the lattice lasers. Finally, spin dynamics is also driven at some point by tunneling processes. However in the Mott regime tunneling-assisted superexchange processes are happening at longer time scales and remain irrelevant for the current measurements.

{\it  Correlation dynamics}
During the evolution under $\hat H_{\rm dd}+\hat{H}_Q$, $\langle\hat S_z \rangle$ and ${\rm{Var}}(\hat S_z)$ are constant, as these two operators commute with $\hat{S}_z$; on the contrary, interactions between spins lead to evolution of spin populations, hence of $\Sigma_z$ and $C_z$. In our case, as spins are orthogonal to the magnetic field, $ \langle\hat S_z \rangle=0$ and Eq.(\ref{Sigmaz}) reads $\Sigma_z=N\sum_{m_s}p_{m_s}m_s^2$; besides ${\rm{Var}}(\hat S_z)=\frac{3}{2}N$ as the initial state is a coherent spin state. The short-time evolution is obtained by perturbation theory  \cite{Lepoutre2019}, leading to $ C_z\approx -\frac{45N}{8}t^2(3V_{\rm eff}^2-4B_Q)$, where $V_{\rm eff}^2=\sum_{i,j\neq i}^NV_{ij}^2/(2N)$, $V_{\rm eff}\simeq h \times4.3$ Hz. At longer times, we can numerically simulate the dynamics via a semiclassical phase space method known as the generalized discrete truncated Wigner approximation (GDTWA) \cite{Zhu2019}, which was previously shown to capture quantitatively the spin population dynamics of this system \cite{Lepoutre2019}.

We also provide a theoretical estimate of the expected correlation at long time  assuming the Eigenstate Thermalization Hypothesis \cite{Alessio2016, Kaufman2016}. In this case, due to the build up of quantum correlations, local observables at long time can  be described  by a thermal density matrix with additional Lagrange multipliers that account for conserved quantities. A high-temperature $T$ series expansion valid for our system \cite{Lepoutre2019,FutureWork} leads to $C_{z}(t\to \infty)=\left(-\frac{5}{2}+12 \beta B_{\rm Q}\right)N$, with $\beta\equiv\frac{1}{k_B T}=\frac{5 B_{\rm Q}+9\bar{V}}{48 V_{\rm eff}^2+ 24 B_{\rm Q}^2}$, and $\bar{V} =\frac{ 1}{N}\sum_{i> j}^N V_{ij}\simeq h \times-0.6$ Hz.

Experimentally, the quantities of interest are the total number of atoms, $N(t)$, and the fractional spin populations
$p_{m_s}(t)$. While the fluctuations in $N(t)$ from shot to shot (with a standard deviation of about $10\%$) yield large extra fluctuations on the measured absolute spin populations $N_{m_s}=N p_{m_s}$,  this extra source of noise is cancelled when dealing with fractional populations.
For measuring the total atom number, we use absorption imaging of the BEC. We checked that the loading in the optical lattice does not lead to losses and therefore  $N(t=0)$ is  equal to the atom number in the BEC. We estimate the accuracy of this measurement equal to $10\%$.

To measure $p_{m_s}(t)$, we spatially separate the $7$ spin components during a time of flight of $14$ ms, using a Stern Gerlach (SG) technique. We use fluorescence imaging to count atoms: it brings equal detectivity of spin components, and makes the use of Electron Multiplying (EM) on the CCD camera advantageous for signal to noise ratio (see  \cite{SuppMat} for details). Atoms are excited by a saturating laser set at 425 nm (with a transition rate $\Gamma=2\pi\times5\times10^6$ Hz) during typically $500$ $\mu$s. The magnetic field $B_0$ is reduced to a small value ($g\mu_BB_0\ll h \Gamma$) to ensure that the fluorescence rates of the 7 spin components are almost equal. We use a ``delta-kick" stage \cite{deltakick} at the very beginning of the time of flight: it consists of a short 0.5 ms pulse of an intense IR laser along the separation axis of the SG  that applies a force on the atoms and helps reducing velocity dispersion. We fine-tune  the frequency of the laser exciting the atoms and the amplitude of all three components of the magnetic field during the fluorescence stage. The obtained regular shape of clouds  (see Fig.\ref{fig1}(b)) favors efficient fitting.

\begin{figure}
\centering
\includegraphics[width= 3.3 in]{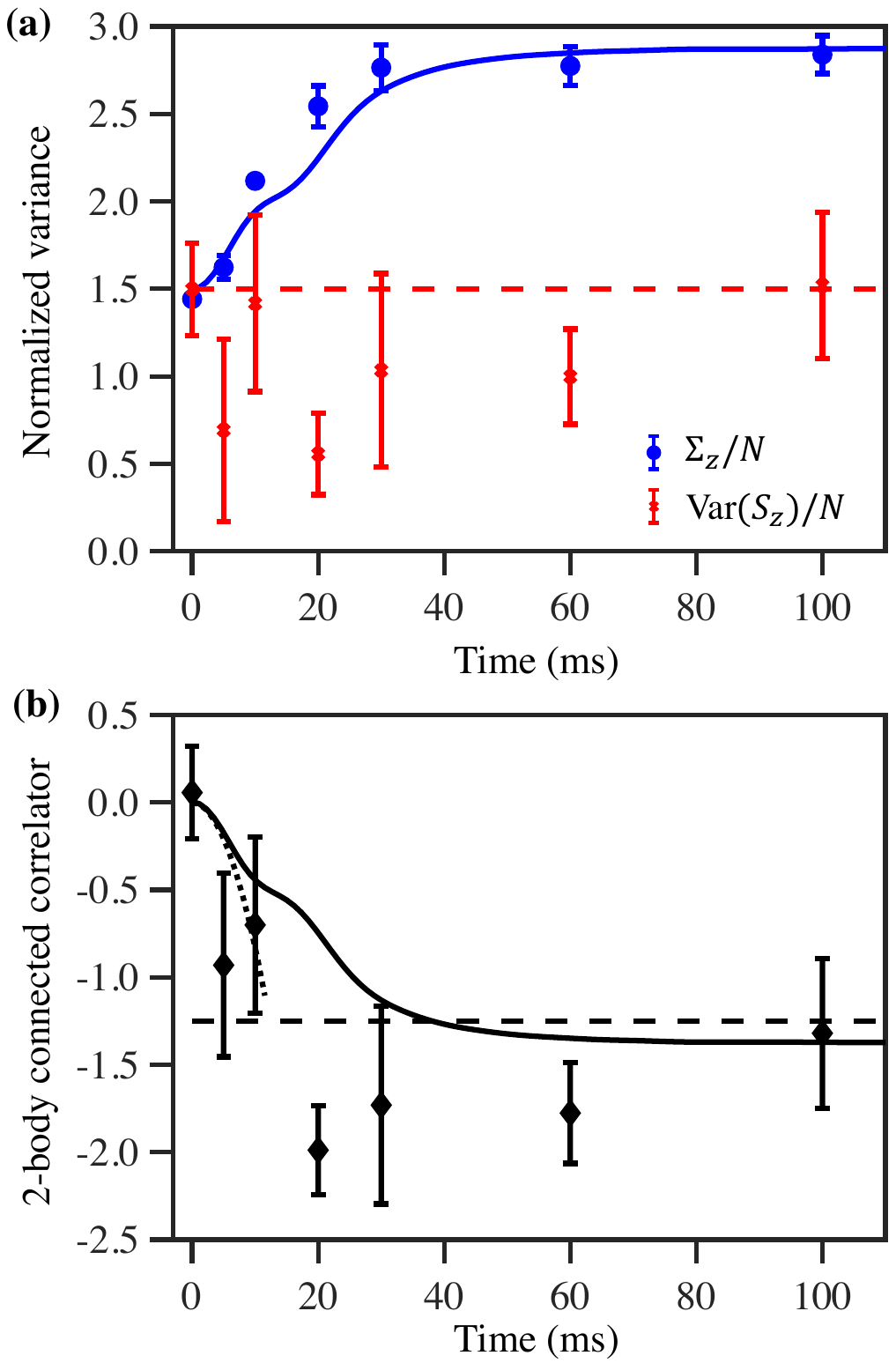}
\caption{(a) Symbols are experimental values for the two contributions to the correlator $C_z$ (see eq.(\ref{Cz})) normalized to atom number. Full line is results of simulations for $\Sigma_z$, while the dashed line shows the expected value for ${\rm{Var}}(\hat S_z)$  (see text). (b) Experimental value of the correlator $C_z$ normalized to the atom number (symbols), with comparison to simulations (full line), and short time expansion (dotted line). The dashed line corresponds to the calculated value in the quantum thermalized state.  Error bars evaluated from statistics correspond to two standard deviations.}
\label{fig3}
\end{figure}

By fitting of the atomic clouds with a Gaussian function we obtain the values of the number of counts $C_{m_s}$ detected for every spin components $m_s$ which  set  the value of $p_{m_s}=C_{m_s}/\sum_{m_s}C_{m_s}$. During the dynamics, $N(t)$ is deduced by multiplying $N(0)$ by the ratio of the total number of counts at $t$ and at $t=0$.

As explained above, ${\rm{Var}}(\hat S_z)(t)$ is expected to be equal to $\frac{3}{2}N(t)$ for a dipolar system without losses. But we do not assume that this equality holds, we measure  ${\rm{Var}}(\hat S_z)(t)$ by  thorough investigations of all different sources of noise. In practice, we measure the variance of the normalized  magnetization of the sample, $M_z=\sum_{m_s}p_{m_s}m_s$, $-3\leq M_z\leq3$, from 3 to 5 sets of 40 pictures. In absence of noise, ${\rm{Var}}(\hat S_z)=N\times{\rm{Var}}(M_z)$.
But at $t=0$, we obtain $N(0)\times{\rm{Var}}(M_z)\simeq 2\times \frac{3}{2}$, which shows that noise processes come into play in our measurement of $M_z$: a proper determination of ${{\rm Var}(\hat S_z)}$ requires to evaluate their contribution independently.

The noises on $M_z$ originate from fluctuations in the preparation angle $\theta$, in the detection process (due to the Poissonian nature of light), and in the evaluation of counts on the camera (related to error in the fitting procedure). We denote their  respective contribution to the  standard deviation on $M_z$ as $\sigma_{{\rm{prep}}}$,  $\sigma_{{\rm{shot noise}}}$ and  $\sigma_{{\rm{fit}}}$. These  different noises are statistically  independent, so that
\begin{equation}
{\rm{Var}}(M_z(t))=\frac{{\rm{Var}}(\hat S_z(t))}{N^2(t)}+\sigma_{{\rm{shot  noise}}}^2(t)+\sigma_{{\rm{fit}}}^2(t)+\sigma_{{\rm{prep}}}^2 \label{sumsigmaMz}
\end{equation} from which we  derive ${\rm{Var}}(\hat S_z)(t)$ at  any time $t$.

From a first principle calculation  we can  determine  $\sigma_{{\rm{shot noise}}}$  from the average counts $C_{m_s}$ and camera parameters, see details in  \cite{SuppMat}. Similarly, $\sigma_{{\rm{fit}}}$ is well evaluated from data analysis. We use measurements at $t=0$ to evaluate the last contribution, $\sigma_{{\rm{prep}}}$. Indeed, the initial sample corresponds to  an uncorrelated spin coherent state, for which ${\rm{Var}}(\hat S_z)=\frac{3}{2}N(0)$ is guaranteed. The conservation of magnetization during the whole spin dynamics ensures that $\sigma_{{\rm{prep}}}$ is constant, as discussed in \cite{SuppMat}; we stress that the contribution of the preparation noise becomes negligible at long time, see Fig. \ref{fig2}.

The noise contributions as dynamics proceeds are shown in Fig. \ref{fig2}, and compared to the one of atomic projection noise, $\sigma_{\rm{SQN}}=\sqrt{\frac{3}{2N(t)}}$. We obtain $\sigma_{{\rm{prep}}}=0.008\simeq0.7\sigma_{\rm{SQN}}$. As $\sigma_{{\rm{prep}}}$ scales like $s \times g_L$ and is independent of $N_0$, while $\sigma_{\rm{SQN}}$ scales like $\sqrt{\frac{s}{N(0)}}$ we stress the difficulty to get such a low value with large $N(0)$, large spin $s=3$ and large Land\'e factor $g_L=2$. Figure  \ref{fig2} shows that $\sigma_{{\rm{shot noise}}}$ scales as $1/\sqrt{N(t)}$ as expected, and that $\sigma_{{\rm{fit}}}$ has about the same scaling.

We show our measurements of ${\rm{Var}}(\hat S_z)(t)/N(t)$ in Fig. \ref{fig3}(a). The scatter of the data points around $\frac{3}{2}$ is comparable to the average error bars for the different points. This indicates that deviations compared to $\frac{3}{2}$ are not statistically significant throughout the curve. This arises because the number of shots taken to estimate ${\rm{Var}}(\hat S_z)$ at each time ($\simeq 150$) is not large enough for the noise associated with finite data sampling to be negligible - a well-known difficulty when estimating correlations from noise analysis.

As we measure a substantial growth for $\Sigma_z(t)/N(t)$, we can assert that the correlator $C_z(t)$ significantly differs from zero for $t>20$ ms, as directly shown in Fig.\ref{fig3}(b). Fig. \ref{fig3}(a) shows a good quantitative agreement between the measured $\Sigma_z(t)/N(t)$ and predictions from our GDTWA simulations assuming only singlons, while Fig. \ref{fig3}(b) shows qualitative agreement for the measured $C_z(t)$ with our short time expansion. The value of the quadratic term $B_Q$ in simulations, $B_Q=h\times -5.1$ Hz, is inferred from population analysis during the whole dynamics \cite{SuppMat}; it leads to $\frac{C_{z }}{N}(t\to \infty)\simeq -1.3$, in good agreement with the data.

Our measurements  thus quantify the amount of two-body correlations in the expected highly correlated state reached at long time. Assuming translational invariance and isotropic correlations decaying exponentially with a correlation length $\xi$, the measured $C_z$ and $\Sigma_z$ at long time can be related to the onset of correlations with $\xi \approx 0.3$ (in units of the lattice spacing) \cite{SuppMat}. This estimate represents a lower bound to the actual correlation length (assuming concentration of correlations at short distance); its rather small value is nonetheless compatible with the scenario of thermalization at high-temperature.

We now discuss the influence of losses. As dipolar spin exchange dynamics proceeds, doublons can get correlated with surrounding singlons, resulting in a modification of singlons fluctuations. Therefore, quantum fluctuations of the sample, and consequently its quantum correlations, may differ from the singlon-only case. Taking losses into account rigorously is difficult and would require new theoretical models to be developed, which is beyond the scope of this paper. We discuss simple arguments in  \cite{SuppMat} to estimate the contribution of losses on ${\rm{Var}}(\hat S_z)(t)$, and predict small corrections, at the 10 percent level. An improved experimental resolution would be necessary to show deviation from a fully unitary  system.

In conclusion, we have measured the growth of correlations in a large ensemble of interacting spins by analyzing the fluctuations of the collective magnetization. This achievement illustrates the new possibilities offered by $s>1/2$ species, and represents an important step towards understanding the  complex quantum many-body dynamics in   state-of-the-art simulators  of quantum magnetism.

\vspace{1cm}

Acknowledgements: We  acknowledge careful review of this manuscript and  useful  comments  from Thomas Bilitewski and Lindsay Sonderhouse.
The Villetaneuse group acknowledges financial support from CNRS, Conseil R\'egional d’Ile-de-France under Sirteq Agency, Agence Nationale de la Recherche (project ANR-18-CE47-0004), and QuantERA ERA-NET (MAQS project). A.M.R is supported by the AFOSR grant FA9550-18-1-0319, AFOSR MURI, by  the DARPA DRINQs grant, the ARO single investigator award W911NF-19-1-0210,   the NSF PHY1820885, NSF JILA-PFC PHY-1734006 grants, and by NIST.

\vspace{10cm}
\section{Supplemental material}

\subsection{Simulations supporting the small contribution of inhomogeneities}

Here  we  quantify    the corrections induced by   the position dependent magnetic field by explicitly  accounting  for it  in  GDTWA simulations. We use  numerical GDTWA simulations to justify the validity of the   homogeneous approximation used in the main text. Explicitly we add  a term of the form $\sum_{\bf i} B_{\bf i} {\hat s}_z^{\bf i}$ to our simulations with $B_{\bf i}=-4.4i_x-5.1i_y+8.7 i_z$Hz, and  ${\bf i}=(i_x,i_y,i_z)$ are integers labeling the 3D lattice coordinates. The chosen  homogeneity  is consistent with  the expected value from experiment.  As a result of the spatially varying field,  the spins precess at different rate and $\langle \hat s_z^i\rangle\neq \langle \hat S^z\rangle/N$.

\begin{figure}
\centering
  \includegraphics[width= 3.3 in]{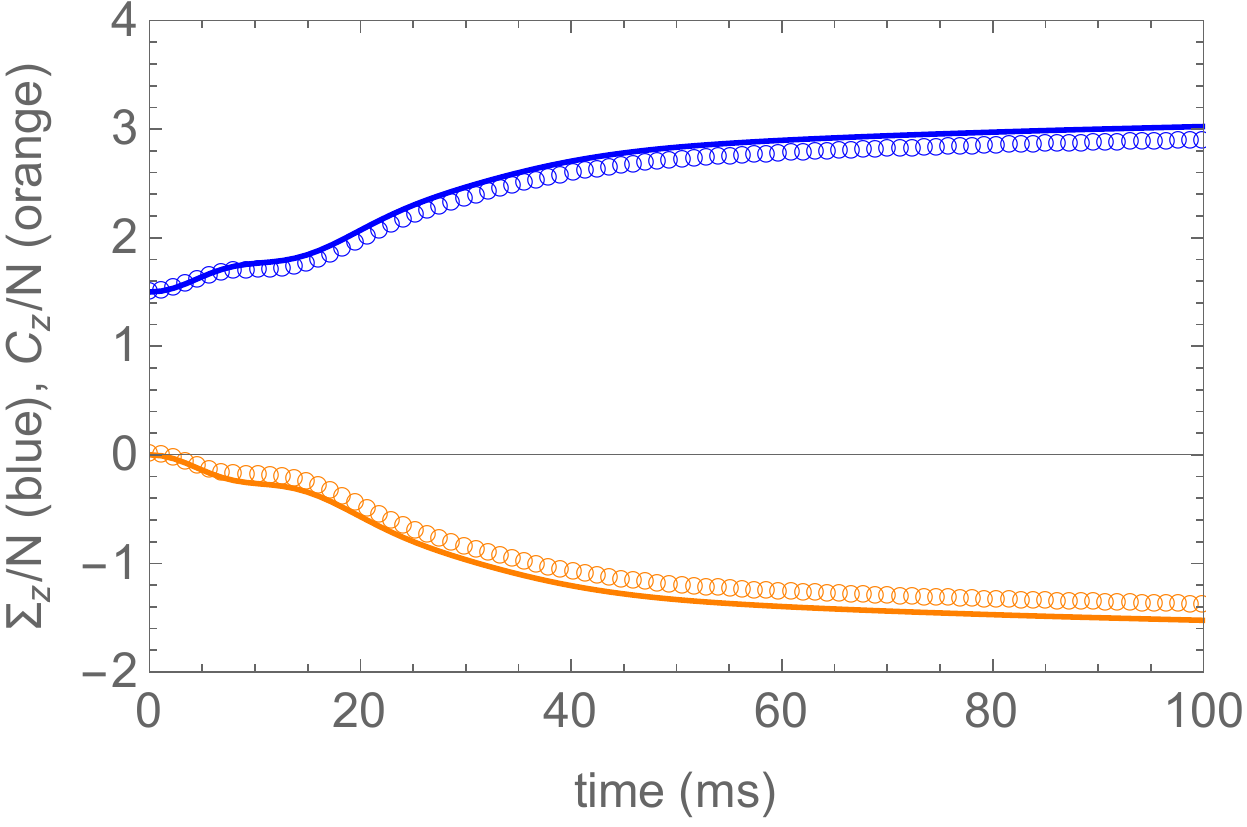}
   \caption{Effect of inhomogeneity. In the presence of magnetic field gradients, comparing the numerical results of $C_z/N$ (orange) and $\Sigma_z/N$ (blue) assuming homogeneous $\langle \hat s_z^i\rangle=\sum_{m_s}m_sp_{m_s}$ (solid lines) and without such assumptions (circles) shows that for our experimental parameters the effect of magnetic field gradients on the local magnetization is sufficiently small to be ignored in the analysis of the collective spin variance.  }\label{fig:Bgrad}
\end{figure}

In Fig. \ref{fig:Bgrad}, we  show the average correlators $C_z$ and $\Sigma_z$  generated by the inhomogeneity during the quantum dynamics. Whilethere is a nonzero net  spin projection along $z$, which causes the difference between results obtained assuming homogeneous $\langle \hat s_z^i\rangle$ (solid lines) and those with actual  $\langle \hat s_z^i\rangle$  (circles), it is much smaller than the value of the  relevant correlators.

\vspace{0.5cm}

\subsection{Calculation of $\Sigma_z$ parameter for doubly occupied sites}\label{Sigma}

Here we focus on the case  of doubly occupied sites  and  explain ways  to   determine    the relevant  correlations considered in this work  in terms of observables that can be  properly measured in the experiment. Particularly, we want to find out how to connect the intra-spin correla
tions $\Sigma_z$ to spin populations.

For an ensemble of $N_d$ doublons, one atom can become correlated with the other in the same site in a different way than with atoms in different sites, and the correlations can be split accordingly:
\begin{align}
\langle (\hat S_z)^2\rangle&=\sum_i\langle (\hat s^{i}_z)^2\rangle+\sum_{i,l\neq i}\langle\hat s^i_z\hat s^l_z\rangle\nonumber\\
&=\sum_i\langle (\hat s^{i1}_z+\hat s^{i2}_z)^2\rangle+\sum_{i,l\neq i}\langle\hat s^i_z\hat s^l_z\rangle={\Sigma}_{aa}+{\Sigma}_{ab}+\mathcal{C}_{z},  \label{eq:corr1}
\end{align}
where $i,l$ label different sites, $\hat s^{i}_z=\hat s^{i1}_z+\hat s^{i2}_z$ accounting for two atoms in site $i$,  $\Sigma_{aa}=\sum_i\langle (\hat s^{i1}_z)^2+(\hat s^{i2}_z)^2\rangle$, and $\Sigma_{ab}=2\sum_i\langle \hat s^{i1}_z\hat s^{i2}_z\rangle$.

Given that the doublons  start fully polarized $S=6$ at the initial time, then the main assumption is that there is a large energy gap opened by the contact interactions that prevents demagnetization of the local spin (see  Ref. \cite{petrapra2019}).  Under this assumption  we can  obtain the dynamics of populations on the  thirteen  different  components, $N_M$ with $M\to \{-6,-5,\dots, 5,6\}$, of an array of $S=6$  spins, and then relate the  measured spin populations $p_{m_s}$  with $m_s\to \{-3,-2,  \dots, 2,3\}$ to   $N_M$:
\begin{eqnarray}
   2N_d  p_{m_s} &=&2\sum_{M=-6}^{6} [C^{6M}_{3m3M-m}]^2N_M,\label{eq:pms6}
\end{eqnarray}Here,  $C^{6M}_{3m3M-m}$ is a Clebsch-Gordon coefficient. To avoid confusion, in the following, we will use $s$ and $m$ symbols for spin-3 operators, and $S$, $M$ for spin-6 operators.

Since there are two atoms in the same site, there are  correlations of the form $\Sigma_{aa}^i\equiv \langle(\hat s^{i1}_z)^2+(\hat s^{i2}_z)^2\rangle$ for each site $i$. At a given lattice site we can write the   doublon's wavefunction as $\left|\psi\right>=\sum_Mc_M\left|M\right>$. Using it, we can obtain the following expression  for the intra-spin correlations at each site:

  \begin{eqnarray}
     \nonumber  \Sigma_{aa}^i&=&\langle (\hat s^{i1}_z)^2+(\hat s^{i2}_z)^2\rangle\\ \nonumber
     &=&\sum_{M,M'}c_M^*c_{M'}\langle SM|(\hat s^{i1}_z)^2 +(\hat s^{i2}_z)^2|SM'\rangle\\ \nonumber
     &=&\sum_{M,M'}c_M^*c_{M'}\sum_{m1,m2,m1',m2'}C^{SM}_{jm1jm2}\langle m_1m_2|(\hat s^{i1}_z)^2\\ \nonumber
     &+& (\hat s^{i2}_z)^2|m_1'm_2'\rangle C^{SM'}_{jm1'jm2'}\\ \nonumber
     &=&\sum_{M,M'}c_M^*c_{M'}\sum_{m1,m2}C^{SM}_{jm1jm2} (m_1^2+m_2^2)C^{SM'}_{jm1jm2}\\ \nonumber
     &=&\sum_{m1,m2}\sum_{M}(m_1^2+m_2^2)|c_M|^2[C^{SM}_{jm1jm2}]^2\\ \nonumber
     &=&\sum_{m1}m_1^2\sum_{m2}\sum_M N^i_M [C^{SM}_{jm1jm2}]^2\\ \nonumber
     &+&\sum_{m2}m_2^2\sum_{m1}\sum_M N^i_M [C^{SM}_{jm1jm2}]^2\\ \nonumber
     &=&2\sum_{m1}m_1^2\sum_{m2}\sum_M N^i_M [C^{SM}_{jm1jm2}]^2\\
     &=&\sum_{m_s} {m_s}^2 p^i_{m_s}\label{eq:measS6},
 \end{eqnarray}
 and $\sum_{i}\Sigma_{aa}^i/N_d=\sum_{m_s} m_s^2 p_{m_s}$. Therefore even for  the $S=6$ manifold, such a relation remains valid, as in the $S=3$ case. For a homogeneous system we thus again obtain $\Sigma_z=N\sum_{m_s}p_{m_s}m_s^2-\langle \hat S_z\rangle^2/N$ even when there are doublons present. With this equation we can obtain the spin-spin correlations $C_z=\mathcal{C}_z+\Sigma_{ab}$ from experimental population measurements even  in the presence of doublons.
While we don't separately consider $\Sigma_{ab}$ and  $\mathcal{C}_z$ in the main text,  in Sec. \ref{comp}, we provide their dynamics from simulations and show that they can  be readily obtained from  experimental data with the knowledge of $\Sigma_{aa}$.

\vspace{0.5cm}

\subsection{Details on Imaging System}
We collect fluorescence light with a 2 inches diameter, 20 cm focal length achromat. The collection efficiency of fluorescence light is $\simeq4\times10^{-3}$. The imaging system after the lens collecting fluorescence, made of three other achromat lenses ensuring a magnification of 1.4, allows to match the size of the full image (the 7 atomic clouds) with the full size of the CCD chip of our EMCCD camera. The camera is cooled at $-90^{\circ}C$, which makes black-body radiation negligible. The quantum efficiency of detection is 0.82 at 425 nm. We use an EM average gain of 24 (which we measure, see section below), and a binning equal to 2 (the counts of 4 adjacent pixels are added together).  Optical shielding all over the imaging path leads to 0.7 photon of stray light per pixel in average.

\subsection{Derivation of the shot noise contribution to the fluctuations of magnetization}
To account for all the possible sources of fluctuations, one needs to consider the physical process that is at play  to estimate the number of atoms, i.e. fluorescence imaging. Atoms emit light, which is an inherently random process characterized by a Poisson distribution. These photons are collected by a camera with quantum efficiency $\eta$, and then the signal is first amplified by a gain $G_1$ referred usually as pre-gain (this first gain is in fact typically a compression $G_1<1$ to optimize the camera well depth to the dynamical range of the digital to analog converter), and then amplified again using an electron amplifier of gain $G_2$, and finally collected.

Let us define $N_{\nu}$ as the photon number that impinges on the camera. This is a fluctuating variable characterized by by a variance $\Delta N_{\nu} = <N_{\nu}>$. We define $N_{e}$ the electronic signal created by the photons arriving at one pixel. One photon creates one electron with a quantum efficiency $\eta$. It is important to remember that this process is stochastic and should be treated as such. There are in fact two independent fluctuating processes: the Poisson fluctuations of light, and the excitation of one electron by one photon in a camera pixel. The variance associated with both these processes need to be calculated, and added in quadrature (since the fluctuations of these process are independent).

For exactly $N_{\nu}$ photons impinging on a camera pixel, the non-unity quantum efficiency $\eta$ of the detector results in a variance
\begin{equation}
    \Delta N_e^{(1)} = (\eta - \eta^2) N_{\nu}.
\end{equation}
In addition to this variance, we need to consider the variance of the photon number, when neglecting the stochastic nature of the detector:
\begin{equation}
    \Delta N_e^{(2)} = \eta^2 \Delta N_{\nu}
\end{equation}
Therefore the total variance in the electron signal is:
\begin{equation}
    \Delta N_e =  \Delta N_e^{(1)} + \Delta N_e^{(2)} =\eta <N_{\nu}> = \eta \Delta N_{\nu}
\end{equation}
We thus simply deduce that $\Delta N_e / <N_e> = \Delta N_{\nu}/ <N_{\nu}>=1$, independent of quantum efficiency. Note that $\sqrt{\Delta N_e} / <N_e> = \frac{1}{\eta} \sqrt{\Delta N_{\nu}} / N_{\nu}$, which indeed shows that the signal to noise may be degraded when $\eta <1$.

We consider now the effect of a deterministic gain $G$. We simply have $S= G <N_e>$ and $\sigma_S = G \sigma_{N_e}$, with $\sigma_{N_e}=\sqrt{\Delta N_e} $. Therefore $\sigma_S / S = \sigma_{N_e}/<N_e> $. The signal to noise is not modified by the gain. Keeping in mind that $\Delta N_e / <N_e> =1$, we finally find:

\begin{equation}
    \sigma_S = \sqrt{G} \sqrt{S}
\end{equation}
As a consequence, when obtaining a signal $S$ on the intensified camera, there is an associated fundamental quantum noise with standard deviation $\sqrt{G S} $. This treatment applies to the pre-gain of the camera, assumed deterministic, which implies that 1 photon gives $G_1$ electrons.

For a camera with an electron multiplying process, one has to consider the stochastic nature of the gain process. For a large number of amplification stages, and a large value of the EM gain $G_2$, this leads to an extra factor $\sqrt{2}$ in the standard deviation, see formula (8) of reference \cite{Robbins2003}:
\begin{equation}
    \sigma_S = \sqrt{2G} \sqrt{S}
    \label{GainEM}
\end{equation}

Taking into account the two gains $G_1$ and $G_2$, we obtain finally:
\begin{equation}
    \sigma_S = \sqrt{2G_1G_2} \sqrt{S}
    \label{GainTot}
\end{equation}
This shows that the whole amplification line needs to be well known in order to predict accurately the contribution of the shot-noise. For our camera settings, $G_1=\frac{1}{4.01}$ is expected from manufacturer data. We optimized the EM gain $G_2$ for our experiment, with a corresponding expected value $G_2=30$. As EM gains are known to be sensitive to aging, we made the following measurements to infer a reliable value of the overall experimental gain $G_{exp}=G_1G_2$.

First, we illuminate the camera with a laser and compare the average number of counts for two camera settings: 1=conventional, i.e. no EM gain, with an expected value of $G_1=3$; 2=with EM gain. The comparison between the average number of counts in the two experiments leads to a value of the average EM gain $G_2=24\pm1$. This simple procedure does not validate the value of $G_1$ in the EM mode.

\begin{figure}
\centering
  \includegraphics[width=0.48\textwidth]{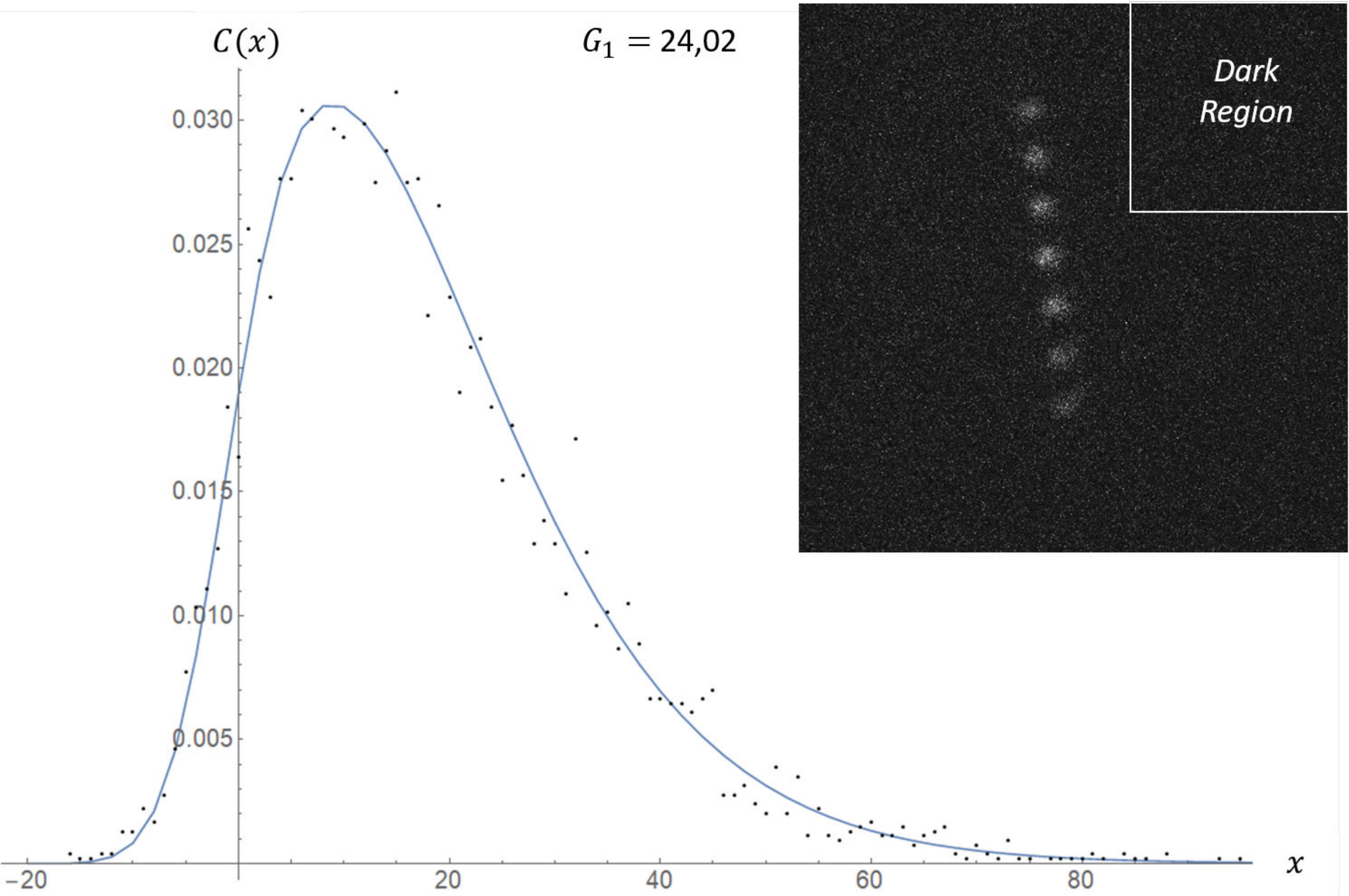}
   \caption{Black points : experimental distribution of the number of counts per pixel. Blue line : Fit of the data. This single parameter fit directly gives the value of $G_2=24.02$ for the image considered here   }\label{fitEM}
\end{figure}

The second method gives a value of the effective gain for each single experimental shot. It relies on the analysis of dimly illuminated regions in each of these shots. From these so called \textit{dark regions} which extend over thousands of pixels ($\geq 5000$), one can extract the probability distribution of the number of counts per pixel $\mathcal{C}(x)$. Fitting this empirical distribution with the appropriate model -see fig.\ref{fitEM} - and assuming $G_1 =\frac{1}{4}$ (as per manufacturer's specifications), one can conclude to the value of the $G_2$ gain. The model is summarized within the following Cauchy product
\begin{equation}\label{convolution}
    \mathcal{C}(x)=\sum_{i=-i_{max}}^{i_{max}} \mathcal{N}_{0,\sigma}(i)\times\sum_{j=-2}^1\mathcal{F}(4(x-i)+j)
\end{equation}
Where:
\begin{itemize}
\item $\mathcal{F}(x)=p_0\delta_{x}+\sum_{n_p} p_{n_p}\frac{x^{n_p-1}e^{-\frac{x}{g}}}{g^{n_p} (n_p-1)!}$ describes the amplification process of Poisson distributed input charges ($p_{n_p}=\lambda^{n_p}\frac{ e^{-\lambda}}{n_p!}$)
\item $\mathcal{N}_{0,\sigma}$ accounts for the read-out noise.
\end{itemize}
While this model seemingly involves many parameters ($G_2,\ \lambda,\ \sigma $), all of these can actually be expressed in terms of $G_2$ through the following equations
\begin{equation}
\begin{split}
    \frac{\lambda G_2}{4}&\simeq \text{mean of empirical data }\\
    \frac{2 G_2^2\lambda}{4^2}+\sigma^2&\simeq \text{ variance of empirical data }\\
\end{split}
\end{equation}

In practice, the mean gain over all experimental series is $\overline{G_2}=23.46$. The standard deviation of the mean gain $\overline{G_2}$ for all experimental series is 0.48. The standard deviation of $G_2$ within a single experimental series is $0.8$. The typical fit standard error is $0.22$. These values are in very good agreement with the results of the first method.

For this value of $G_2$, and our camera having 590 amplification stages, validity of eq.(\ref{GainEM}), hence of eq.(\ref{GainTot}), is expected to be better than $99\%$ according to formula (8) in \cite{Robbins2003}.

Finally, we give the expression of the contribution of the shot noise on the standard deviation of normalized  magnetization $M_z$ in our system. With  $S_{m_s}$ the signal for the spin components $m_s$, $M_z=\sum_{m_s} m_s S_{m_S}/\sum_{m_s} S_{m_S}$, one gets, assuming independent noises:
\begin{equation}
    \sigma_{SN} = \frac{\sqrt{2 G_{exp}\sum_{m_s} m_s^2 S_{m_S}}}{\sum_{m_s} S_{m_S}}
\end{equation}
The corresponding variance is proportional to the total number of atoms $N$, and inversely proportional to the photon collection efficiency.

The shot noise has a significant impact on the observed magnetization fluctuations, as described in the main paper. In our analysis, we thus subtract from the experimental variance the sum of the estimated variance due to shot noise and of the estimated variance due to fit uncertainty. The latter is then estimated assuming that the noise on each pixel of the image is independent of the signal at this pixel.

As shot noise can be non-negligible and is signal-dependent, it is not a priori justified to assume a signal-independent noise to deduce the fit-noise. We therefore have also estimated the combined effect of the shot noise and other signal-independent noises (such as read noise) in the following manner: we estimate the variance of the fit, using a $\chi^2$ that is now normalized to the noise on each pixel. This latter noise is obtained by adding in quadrature the signal-independent noise (obtained by statistics on the parts of the camera where the signal is negligible) and the estimated shot noise per pixel (see above). We find that both methods give very similar results, which validate the approach that we follow in the main part of the paper.
\vspace{0.5cm}

\subsection{Details of numerical simulations and determination of the quadratic field value}

Here we provide details regarding our numerical simulations, as well as the determination of the quadratic light shift $B_Q$ induced by the lattice lasers, which is the only free parameter used in our numerical simulations in the main text. To examine the population and correlation dynamics for large system sizes, we make use of the generalized discrete truncated Wigner approximation (GDTWA), previously introduced in \cite{Zhu2019}.

\begin{figure}
\centering
  \includegraphics[width=0.48\textwidth]{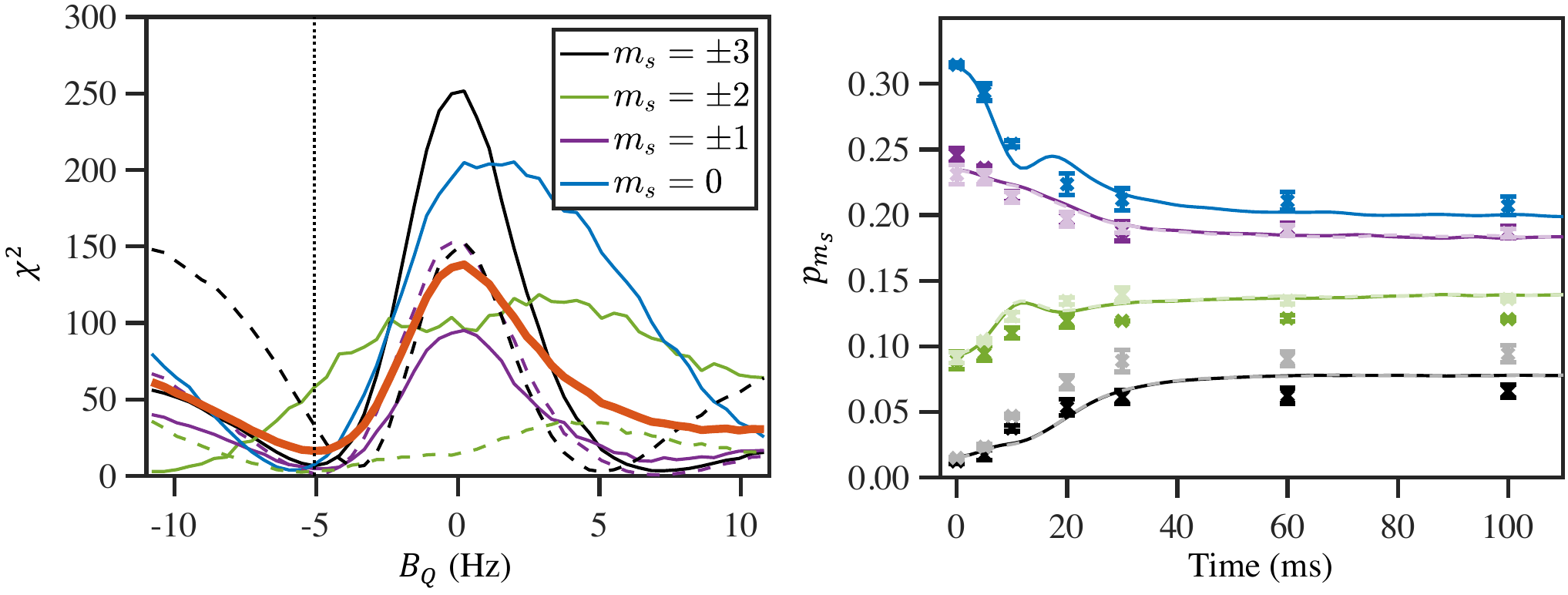}
   \caption{Determination of the quadratic light shift from the experimental spin populations. (Left) Value of $\chi^2 = \overline{\chi_{m_s}^2}$ (orange line) for various $B_Q$. We also plot $\chi^2_{m_s}$ for each $m_s$ averaged over all available times; dashed (solid) lines denote positive (negative) $m_s$. Vertical dotted line denotes optimal value of $B_Q$ (minimum of $\chi^2$). (Right) GDTWA results for best-fit $B_Q \approx -5.1$ Hz, compared with experimental results for the population dynamics; positive (negative) $m_s$ displayed in lighter (darker) shading.  }\label{BestBQ}
\end{figure}

In order to make a comparison with experimental results, we must first determine the best-fit value of the quadratic light shift, which is known to be present in the experiment. We perform numerical GDTWA simulations of the Hamiltonian in the main text with an added term of the form $B_Q\sum_i (\hat{s}_z^i)^2$. For each value of $B_Q$ that we consider, we compute the quantity $\chi^2_{m_s}(t) = (p_{m_s}^{\left[\mathit{sim}\right]}(t) - p_{m_s}^{\left[\mathit{exp}\right]}(t))^2/\sigma^2_{m_s}(t)$, where $p_{m_s}^{\left[\mathit{sim/exp}\right]}(t)$ denotes the simulated/experimental values of $p_{m_s}(t)$, and $\sigma^2_{m_s}(t)$ corresponds to the experimental error (we assume comparatively negligible GDTWA sampling error). We then average over all $m_s$ and available $t$ to obtain the mean-squared error $\chi^2 = \overline{\chi^2_{m_s}(t)}$. Minimizing this quantity over $B_Q$, we find an optimal value of $B_Q \approx -5.1$ Hz with $\chi^2 = 15.3$. In Fig.~\ref{BestBQ}, we plot $\chi^2$ over the range of $B_Q$ we consider, and compare the resulting dynamics of $p_{m_s}$ for the best-fit $B_Q$. We observe that our simulations provide decent agreement with the experimental results, capturing the relaxation timescales and steady-state values of $p_{m_s}$.

All of our results are obtained for a lattice size of $L_x\times L_y \times L_z = 5\times 3\times 5$ with periodic boundary conditions, which we have verified is enough to produce results for $p_{m_s}(t)$ that are convergent in system size along each dimension, within relevant timescales and experimental error bars.

\vspace{0.5cm}

\subsection{Estimate of the correlation length from the measurement of $C_z$ and $\Sigma_z$ }

In this section we provide an estimate of the characteristic correlation length associated with the dynamical onset of correlations in the system, starting from the measurement of $C_z$ and $\Sigma_z$.

To this scope, let us introduce the spin-spin correlation function $G_z(i,j) = \langle s_z^i s_z^j \rangle - \langle s_z^i \rangle \langle s_z^j\rangle$. When focusing on the long-time regime, in which the system is expected to thermalize, we can in general expect $G_z$ to be short-ranged, namely decaying exponentially with the distance between the sites, with a decay rate given by the correlation length $\xi$. The spatial anisotropy of the dipolar interactions, as well as of the optical lattice used in the experiment, would suggest that there are in fact  several correlation lengths when moving along different lattice directions; nonetheless for simplicity we shall neglect this aspect, and assume that correlations are spatially isotropic.

Moreover, in the same spirit we shall assume that the system is translationally invariant, namely that $G_z(i,j) = G_z(r_{ij})$, where $r_{ij}$ is the distance between the $i$-th and $j$-th spin. $G_z(r)$ can therefore be taken to be $|G(z)|  \approx G_z(0) \exp(-r/\xi)$. The zero-range correlations $G_z(0)$ coincide with the on-site spin fluctuations, namely with $\Sigma_z/N$. On the other hand $C_z$ is the sum of the offsite correlations; given that $C_z$ is negative, it is plausible to assume that the sign of all the correlations for $r>0$ is globally negative. Putting all these assumptions together, we posit the simple functional form $G_z(r>0)  \approx - G_z(0) e^{-r/\xi}$ for the off-site correlations.

By definition of $C_z$ we have that
\begin{equation}
\frac{C_z}{N} = \sum_{\bm r \neq 0} G_z(|\bm r|) \approx \frac{\Sigma_z}{N} \sum_{\bm r \neq 0} e^{-r/\xi}
\end{equation}
from which we deduce the integral relationship between $C_z$, $\Sigma_z$ and $\xi$
\begin{equation}
    \sum_{\bm r \neq 0} e^{-r/\xi} = \frac{|C_z|}{\Sigma_z}~.
    \label{e.xi}
\end{equation}

At long times ($t = 100$ ms) we measure $C_z/N \approx -1.3$  and $\Sigma_z/N \approx 2.8$. Solving Eq.~\eqref{e.xi} for $\xi$ numerically, we obtain the value $\xi = 0.315$ quoted in the main text.

Let us remark that the working assumption $G_z(r>0)  \approx - G_z(0) e^{-r/\xi}$, which attributes the same amplitude to the on-site fluctuations and to the exponential tail of the correlations, is actually assuming a maximum concentration of correlations at short distances. Therefore its use provides in practice a \emph{lower bound} to the correlation length $\xi$. A more general form would have been $G_z(r>0) \approx -A e^{-r/\xi}$ with $A \leq G_z(0)$, leading to a larger $\xi$ estimate. Yet the experimental data at hand do not allow us to estimate $\xi$ and $A$ independently.

\subsection{ Effect of losses on correlations}

To estimate the effect of losses on the measured covariances and correlations, we have developed a simple heuristic statistical model. This model ignores the effect of the dipole-dipole interactions between atoms that is described by the secular Hamiltonian in the paper, but it includes dipolar relaxation. It considers the case of a system where all spin are initially tilted by $\theta = \pi/2$ compared to the magnetic field.

Consider the initial number of single- and double-occupied sites, denoted $S$ and $D$, respectively. The probability that a given atom initially occupies a single-occupied site is then $P_S = S/(S+2D)$, while the probability that it occupies a doubly occupied state is $P_D = 2D/(S+2D)$. Since the magnetic field is sufficiently large, dipolar relaxation only affects doubly-occupied sites; for an atom occupying such a site and with a spin projection $m_i$, the inverse lifetime due to dipolar relaxation is $\Gamma_{m_i}=\sum_{m_s} p_{m_s} \Gamma_{m_i,m_{m_s}}$. Here, $p_{m_s}$ are the fractional populations, which in principle are time-dependent but we will take to equal the populations in the initial state, since dipolar relaxation lead to small dynamics in the fractional populations \cite{kaci}. The $\Gamma_{m_i,m_{m_s}}$ are calculated in the Born approximation \cite{Pasquiou2010cod,kaci}, and are spin-dependent through the angular terms in the spin operators. Thus, the probability to measure a given atom in the state $m_i$ can be written as
\begin{eqnarray}
    \nonumber P_{m_i} = \alpha \frac{p_{m_i}}{S+2D}\times\left[ S + 2D e^{-\Gamma_{m_i}t}\right].
\end{eqnarray}
The time-dependent constant $\alpha$ is defined by the normalization condition $\sum_{m_i} P_{m_i} = 1$.

In the absence of interactions, two atoms on separate lattice sites will decay independently of each other. Since the probability that any two given atoms will occupy the same lattice site is very small, we make the additional assumption that \emph{any} two atoms will decay independently of each other. The measurement of the spin projection $m_i$ over $N$ uncorrelated atoms is then described by a multinomial process with $N$ independent trials and parameters $P_{m_i}$ describing the probability of measurement outcome $m_i$ for each atom. Thus, the variance and covariance are given by
\begin{eqnarray}
{\rm Var}[N_{m_i}]=N P_{m_i}(1-P_{m_i})
\end{eqnarray}
\begin{eqnarray}
{\rm Cov}[N_{m_i},N_{m_j}]= -NP_{m_i}P_{m_j}.
\end{eqnarray}

This allows for an estimate of ${\rm Var}[S_z]$. As expected we find that at $t=0$ (losses have not occurred) and at large $t$ (where all doublons have disappeared), the outcome is identical, and equal to the expected variance $3/2 N$. For intermediate times,  the variance varies by typically less than 5 percent. This indicates that the effect of losses on correlations may safely be neglected in our experiment.

\subsection{Study of the stationarity of the preparation noise through the whole dynamics}

In the case that there are fluctuations in the initial rotation angle $\theta$, the total variances can be obtained as
\begin{eqnarray}
{\rm Var}^{\rm tot}[ O]&=&\int d\theta p(\theta)\langle \hat O^2\rangle_\theta-\int d\theta p(\theta)\langle \hat O\rangle_\theta\int d\theta p(\theta)\langle \hat O\rangle_\theta,\nonumber\\\label{eq:totvar1}
\end{eqnarray}

where $p(\theta)$ is the distribution of the initial rotation angles in the presence of rf noise.

It is straightforward  to see the above is equivalent to

\begin{eqnarray}
\nonumber {\rm Var}^{\rm tot}[ O]&=&\overline{{\rm Var}[O]_\theta}+\overline{(\langle \hat O\rangle_\theta-\overline{\langle \hat O\rangle_\theta})^2}\label{eq:totvar2}\\ \nonumber
&&=\int d\theta p(\theta)[\langle\hat O^2\rangle_\theta-\langle\hat O\rangle^2_\theta]\\
&+&[\int d\theta p(\theta)\langle \hat O\rangle_\theta^2-(\int d\theta p(\theta)\langle \hat O\rangle_\theta)^2],
\end{eqnarray}

where $\overline{O}$ denotes the average over the distribution of $\theta$, and $\langle\hat O\rangle$ denotes the quantum expectation value. That is, we can compute the total variance either using Eq.~(\ref{eq:totvar1}) or directly using Eq.~(\ref{eq:totvar2}). In the following, we will further show some simplified results from Eq.~(\ref{eq:totvar2}).

In the case that $p(\theta)$ is a narrow distribution around a certain angle $\theta_0$, eg.
\begin{eqnarray}
p(\theta)&\propto&e^{-\frac{(\theta-\theta_0)^2}{2\sigma^2_\theta}},
\end{eqnarray}
with $\sigma_\theta\ll1$, we can expand the functions inside the integral $\int d\theta p(\theta)F(\theta)$ as
\begin{eqnarray}
F(\theta)\!&\approx&\!F(\theta_0)+\!\frac{\partial F}{\partial \theta}\bigg|_{\theta_0}\!\!\!(\theta-\theta_0)+\frac{\partial^2 F}{\partial \theta^2}\bigg|_{\theta_0}\!\!\!\frac{(\theta-\theta_0)^2}{2}\!+\!\mathcal{O}(\theta-\theta_0)^3.\nonumber\\
\end{eqnarray}
Substitute this into either of the above equations for the total (co)variances, one obtains
\begin{eqnarray}
{\rm Var}^{\rm tot}[O]&\approx&\int d\theta p(\theta) {\rm Var}[O]_{\theta}+\left[\frac{\partial \langle\hat O\rangle}{\partial\theta}\bigg|_{\theta_0}\right]^2\sigma_\theta^2\\
&\approx&{\rm Var}[O]_{\theta_0}+\frac{1}{2}\frac{\partial^2{\rm Var}[O]}{\partial\theta^2}\bigg|_{\theta_0}\sigma_\theta^2+\left[\frac{\partial \langle\hat O\rangle}{\partial\theta}\bigg|_{\theta_0}\right]^2\sigma_\theta^2,\label{eq:totvarapprox}\nonumber\\
\end{eqnarray}
where ${\rm Var}[O]$ is the quantum variance from an initial state $\theta$, as calculated in the previous section. Note, for the initial state considered in experiment,  the first term  ${\rm Var}[O]_{\theta_0}$ is $\propto N$, and the second term $\propto N\sigma_{\theta}^2$, while the last term in  Eq.~(\ref{eq:totvarapprox}) is $\propto N^2\sigma_{\theta}^2$. So when $N\gg 1$, the second term is negligible compared to the last term, and can be dropped. This means roughly one can estimate the total variance as
\begin{eqnarray}
{\rm Var}^{\rm tot}[O]&\approx&{\rm Var}[O]_{\theta_0}+\left[\frac{\partial \langle\hat O\rangle}{\partial\theta}\bigg|_{\theta_0}\right]^2\sigma_\theta^2,\label{eq:totvarapproxbigN}
\end{eqnarray}
ie., the simple summation of the contribution from two quadratures. To have quantum noise that is significant compared to the technical noise contribution, this also suggests that a small rf noise is needed $\sigma_\theta^2\sim 1/N$.

\vspace{0.5cm}

\subsection{Comparison between intrasite and intersite spin correlations}\label{comp}

\begin{figure}
\centering
  \includegraphics[width=0.4\textwidth]{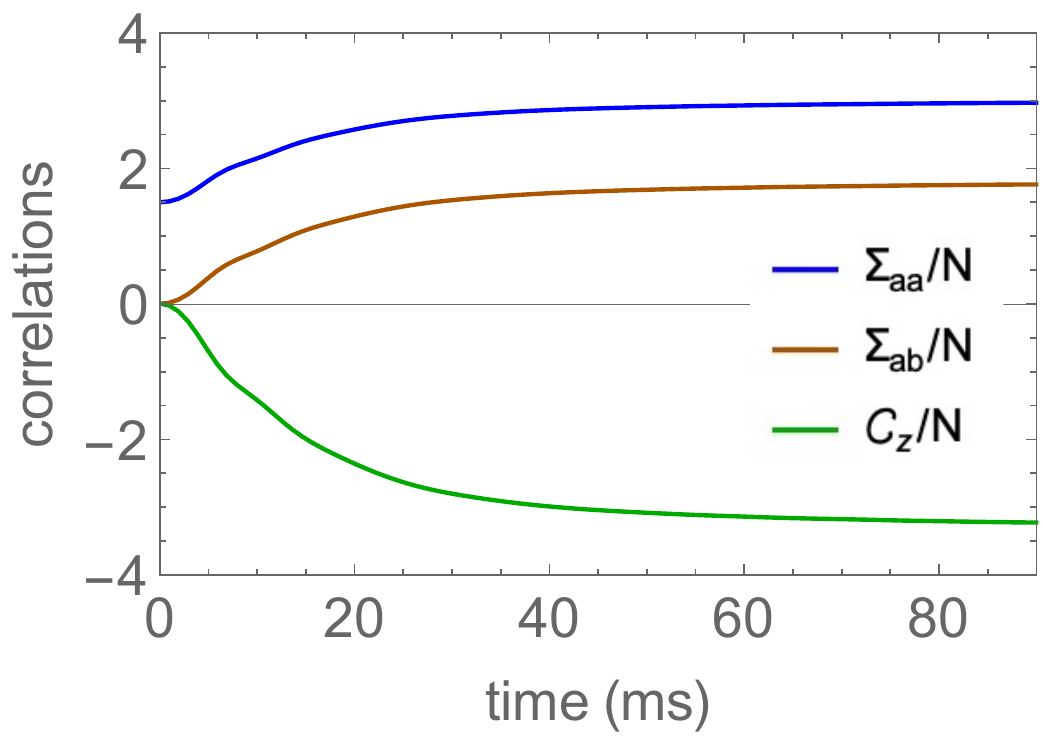}
   \caption{Spin correlations for doublons, $\Sigma_{aa}$ (blue), $\Sigma_{ab}$ (orange), and $\mathcal{C}_z$ (green). Results obtained from GDTWA simulations for a 3D lattice with $N_d=1372$, with $B_Q=-5$Hz and zero magnetic field gradients. The lattice sites are all doubly occupied, with total atom number $N=2N_d$.  }\label{fig:figcorr}
\end{figure}

As discussed in Sec. \ref{Sigma}, in the presence of doublons, in addition to the intra-spin correlations $\Sigma_z$, correlations can build up between atoms in the same site, $\Sigma_{ab}=2\sum_{i}\langle \hat s_z^{i1}\hat s_z^{i2}\rangle$, as well as between different sites, $\mathcal{C}_z=\sum_{i,j\neq i}^{\rm \# of~ sites} \langle \hat s_z^{i}\hat s_z^{j}\rangle$. Describing doublons as $S=6$ particles, we find at short time, the growth of these correlations takes the form
\begin{eqnarray}
\Sigma_{ab}&=&-27t^2(4B_Q\overline{V}-3V_{\rm eff})^2,
\end{eqnarray}
\begin{eqnarray}
\mathcal{C}_{il}&=&\frac{99 t^2}{2}(4B_Q\overline{V}-3V_{\rm eff}^2),
\end{eqnarray}
where $\overline{V}=\sum_{j<k}V_{jk}/2N_d$ and $V_{\rm eff}^2=\sum_{j<k}V_{jk}^2/2N_d$, with the summation running over different lattice sites populated by doublons. That is, the intra-site and inter-site correlations have different growth rates and signs. As shown in Sec. \ref{Sigma}, $\Sigma_{aa}$ can be obtained from experiment with collective measurement. We have verified with numerical simulations that it can be related to $\Sigma_{ab}$ via
\begin{align}
&\Sigma_{ab}\!=\!\frac{6}{5}\left(\Sigma_{aa}(t)-\frac{2}{7+5\cos(2\theta)}\Sigma_{aa}(0)\right),
\end{align}
where $\theta$ is the tipping angle of the initial state, and in this work $\theta=\pi/2$. Then $\mathcal{C}_z$ can be obtained from $\mathcal{C}_z=3N/2-\Sigma_{aa}-\Sigma_{ab}$.

In Fig. \ref{fig:figcorr} we use GDTWA to calculate the dynamics of different correlations up to a long timescale relevant for experiment, which shows that atoms in different sites become significantly anti-correlated under the dipolar interactions in our system.

\vspace{3cm}

\bibliography{biblioCorrelArxiv}

\end{document}